\documentclass[11pt,twocolumn]{article}
\usepackage{graphicx}
\usepackage{amsmath}
\usepackage{bm}

\begin{document}

\title{\textsf{\textbf{Revisiting 2x2 matrix optics: Complex vectors, Fermion combinatorics, and Lagrange invariants}}}
\author{Quirino M. Sugon Jr.* and Daniel J. McNamara
\smallskip\\
\small{Ateneo de Manila University, Department of Physics, Loyola Heights, Quezon City, Philippines 1108}\\
\small{*Also at Manila Observatory, Upper Atmosphere Division, Ateneo de Manila University Campus}\\
\small{e-mail: \texttt{qsugon$@$observatory.ph}}}
\date{\small{\today}}
\maketitle

\small
\section*{}\label{Abstract}
\textbf{Abstract.}  
We propose that the height-angle ray vector in matrix optics should be complex, based on a geometric algebra analysis. We also propose that the ray's $2\times 2$ matrix operators should be right-acting, so that the matrix product succession would go with light's left-to-right propagation.  We express the propagation and refraction operators as a sum of a unit matrix and an imaginary matrix proportional to the Fermion creation or annihilation matrix.  In this way, we reduce the products of matrix operators into sums of creation-annihilation product combinations.  We classify ABCD optical systems into four: telescopic, inverse Fourier transforming, Fourier transforming, and imaging.  We show that each of these systems have a corresponding Lagrange theorem expressed in partial derivatives, and that only the telescopic and imaging systems have Lagrange invariants.

\section{Introduction}

\textbf {a.  Complex Vectors.}  In $2\times 2$ matrix optics, a ray is normally described by a column vector as given by Nussbaum and Phillips\cite{NussbaumPhillips_1976_ContemporaryOpticsforScientistsandEngineers_p10}:
\begin{equation}
\label{eq:r is e_1 x + e_2 nalpha intro}
\mathbf r=
\begin{pmatrix}
x\\
n\alpha
\end{pmatrix}
=\mathbf e_1x+\mathbf e_2n\alpha,
\end{equation}
except that we interchanged the coefficients.  This equation is problematic: $\alpha$ is an angle and not a distance.  The Cartesian coordinate system is a system for locating a point in space in terms of distances from a fixed point measured along orthogonal lines.  But in what space do angles live?  In an imaginary vector space?  

Yes.  To see why this is so, let us first recall the orthonormality axioms in geometric algebra\cite{BaylisHuschiltWei_1992_ajp60i9pp788-797_p789}:
\begin{eqnarray}
\label{eq:e_j squared is e_k squared is 1 intro}
\mathbf e_j^2&=&\mathbf e_k^2=1,\\
\label{eq:e_j e_k is -e_k e_j intro}
\mathbf e_j\mathbf e_k&=&-\mathbf e_k\mathbf e_j,\quad j\neq k,
\end{eqnarray}
where $j,k\in\{1,2,3\}$.  That is, the square of each unit vector is unity and that the product of two perpendicular vectors anticommute.  Notice that the multiplicative inverse of a unit vector is itself.

We know from Yariv\cite{Yariv_1989_QuantumElectronics_p106-107} that the paraxial angle $\alpha$ is the slope of the function $x=x(z)$:
\begin{equation}
\label{eq:alpha as dx by dz intro}
\frac{dx}{dz}=\tan\alpha\approx\alpha.
\end{equation}
But if we define $\mathbf z=z\mathbf e_3$ and $\mathbf x=x\mathbf e_1$, then by the rules of geometric algebra,
we have
\begin{equation}
\label{eq:alpha as dx by dz vector intro}
\frac{d\mathbf x}{d\mathbf z}=\mathbf e_3^{-1}\mathbf e_1\frac{dx}{dz}=\mathbf e_3\mathbf e_1\frac{dx}{dz}=i\mathbf e_2\alpha,
\end{equation}
where $i=\mathbf e_1\mathbf e_2\mathbf e_3$ is the unit trivector that behaves like the unit imaginary number\cite{Jancewicz_1988_MultivectorsandCliffordAlgebrainElectrodynamics_p32}.  Therefore, instead of Eq.~(\ref{eq:r is e_1 x + e_2 nalpha intro}), we write
\begin{equation}
\label{eq:r is e_1 x + e_2 inalpha intro}
\hat r=
\begin{pmatrix}
x\\
n\alpha i
\end{pmatrix}
=\mathbf e_1x+i\mathbf e_2n\alpha,
\end{equation}
which is a complex vector like the electromagnetic field $\hat F=\mathbf E+i\mathbf B$\cite{Hestenes_2003_ajp71i2pp104-121_p110}.  Note that we adopted the convention that lengths like height and radius are dimensionless. (Alternatively, we may replace $x$ by $\zeta x$, where $\zeta$ is a unit quantity with dimension of inverse length).  

\textbf{b.  Fermion Combinatorics.}  In most matrix optics texts, the convention is light travelling left to right.  Yet the left-acting propagation and refraction matrix operators used are multiplied from right to left:
\begin{equation}
\label{eq:r' is M product r intro} 
\mathbf r'=\textsf{M}'_N\textsf{M}'_{N-1}\cdots\textsf{M}'_2\textsf{M}'_1\mathbf r.
\end{equation}
So we propose a more logical way: define the matrix operators to be right-acting\cite{SugonFernandezMcNamara_2008_arXiv0811.3680v1_p9}, so that they multiply from left to right, in the same direction of the light's propagation.  That is,
\begin{equation}
\label{eq:r' is r M product intro} 
\mathbf r'=\mathbf r\textsf{M}_1\textsf{M}_2\cdots\textsf{M}_{N-1}\textsf{M}_N.
\end{equation}
Here, the action of the right-acting matrix is defined by the action of its left-acting transpose, as given in Symon\cite{Symon_1971_Mechanics_p408}:
\begin{equation}
\label{eq:M transpose R is R M intro}
\textsf M^T\cdot\mathbf r=\mathbf r\cdot\textsf M.
\end{equation}

Because the ray operators are $2\times 2$ matrices, we may decompose them as a linear combination of single-element, unit matrices, as done by Campbell\cite{Campbell_1994_JOSAA11i22pp618-622} and Harris\cite{Harris_1997_OptVisSciv74i6pp349-366_p357}:
\begin{eqnarray}
\label{eq:M as M_jk e_jk decomposition intro}
\textsf M&=&M_{11}\textsf e_{11}+M_{12}\textsf e_{12}+M_{21}\textsf e_{21}+M_{22}\textsf e_{22},
\end{eqnarray}
where
\begin{eqnarray}
\label{eq:e_11 and e_12 matrix intro}
\textsf e_{11}=
\begin{pmatrix}
1&0\\
0&0
\end{pmatrix}
,&\quad&
\textsf e_{12}=
\begin{pmatrix}
0&1\\
0&0
\end{pmatrix}
,\\
\label{eq:e_21 and e_22 matrix intro}
\textsf e_{21}=
\begin{pmatrix}
0&0\\
1&0
\end{pmatrix}
,&\quad&
\textsf e_{22}=
\begin{pmatrix}
0&0\\
0&1
\end{pmatrix}
\end{eqnarray}
are the Fermion matrices in Sakurai\cite{Sakurai_1967_AdvancedQuantumMechanics_p28} and Le Bellac\cite{LeBellac_1991_QuantumandStatisticalFieldTheory_p404}.  The dyadics $\textsf e_{12}$ and $\textsf e_{21}$ may represent either the creation operator $\hat a^\dagger$ or the annihilation operator $\hat a$, depending on the column matrix representations of $\mathbf e_1$ and $\mathbf e_2$.

For the ray vector $\mathbf r=x\mathbf e_1+n\alpha\mathbf e_2$, the left-acting propagation and refraction matrices are given in Klein and Furtak (transposed matrices)\cite{KleinFurtak_1986_Optics_p152}:
\begin{eqnarray}
\label{eq:T is 1 + e_12 D/n intro}
\textsf T&=&1+\textsf e_{12}D/n,\\
\label{eq:R is 1 - e_21 P intro}
\textsf R&=&1-\textsf e_{21}P,
\end{eqnarray}
where 1 is the unit matrix.  Thus, a lens system becomes a product of propagation and refraction matrices:
\begin{eqnarray}
\label{eq:M is T_n R_n to T_2 R_2 T_1 R_1}
\textsf M&=&\textsf T_n\textsf R_n\cdots\textsf T_2\textsf R_2\textsf T_1\textsf R_1\nonumber\\
&=&\prod_{k=1}^n(1+\textsf e_{12}D_k/n_k)(1+\textsf e_{21}P_k),
\end{eqnarray}

Later, we shall recast these equations for our complex ray vectors and right-acting matrices.  We shall also present new methods for computing the system matrix $\textsf M$.  These methods are based on the Fermion identies satisfied by the four dyadic operators $\textsf e_{11}$, $\textsf e_{12}$, $\textsf e_{21}$, and $\textsf e_{22}$.  In particular, we shall study the allowed combinations of $\textsf e_{12}D_k/n_k$ and $\textsf e_{12}P_{k'}$ and determine the matrix component basis $\textsf e_{kk'}$ of the product chain.

\textbf{c.  Lagrange Invariants.}  The Lagrange theorem or the Smith-Helmholtz relationship\cite{Meyer-Arendt_1984_IntroductiontoClassicalandModernOptics_p55,BornWolf_1964_PrinciplesofOptics_p165} is stated by Welford\cite{Welford_1974_AberrationsoftheOpticalSystem_pp22-23} as
\begin{equation}
\label{eq:n u eta is n' u' eta'}
nu\eta=n'u'\eta',
\end{equation}
where $n$ is refractive index of the input medium, the object height, $u$ is the angle subtended from the object, and $\eta$ is the height of the object; their primed counterparts correspond to those of the image.  The quantity $nu\eta$ is called the Lagrange invariant.  Later, using our $x$ and $n\alpha$ variables, we shall show that we may recast Lagrange's theorem in Eq.~(\ref{eq:n u eta is n' u' eta'}) in terms of partial derivatives:
\begin{equation}
\label{eq:x partial n'alpha' over x partial n alpha is -1 intro}
\frac{x'}{x}\frac{\partial(n'\alpha')}{\partial(n\alpha)}=-1;\quad x,x'=\textrm{constants}.
\end{equation}
The quantity $|x\delta(n\alpha)|$ is the Lagrange invariant. 

\textbf{d.  Outline.}  We shall divide the paper into five sections.  The first section is Introduction.  In the second section, we discuss the algebra of right-acting matrices and their actions on column vectors.  In the third section, we shall introduce the complex ray vector and its right-acting propagation and refraction matrices.  We shall use the properties of Fermion creation-annihilation matrices to compute the system matrices of thin and thick lenses.  In the fourth section, we shall revisit the classification of optical systems: telescopic, Fourier transforming, inverse Fourier transforming, and imaging.  We shall derive their Lagrange theorems and see if we can define their corresponding Lagrange invariants.  We shall also rederive the Moebius transform and the Newton's equation for the imaging system.  The fifth section is Conclusions.

\section{Matrix Algebra}

Let $\mathbf e_1$ and $\mathbf e_2$ be two orthonormal vectors represented as column matrices,
\begin{equation}
\label{eq:e_1 and e_2 as matrix}
\mathbf e_1=
\begin{pmatrix}
1\\
0
\end{pmatrix}
,\qquad
\mathbf e_2=
\begin{pmatrix}
0\\
1
\end{pmatrix}
,
\end{equation}
and let $\textsf e_{11}$, $\textsf e_{12}$, $\textsf e_{21}$, and $\textsf e_{22}$ be the four Fermion matrices in Eqs.~(\ref{eq:e_11 and e_12 matrix intro}) and (\ref{eq:e_21 and e_22 matrix intro}).  The left and right action of the dyadic operators on $\mathbf e_1$ and $\mathbf e_2$ are defined by the following relations\cite{SugonFernandezMcNamara_2008_arXiv0811.3680v1_p8}:
\begin{eqnarray}
\label{eq:e_lambda e_munu right}
\mathbf e_{\lambda}\cdot\textsf e_{\mu\nu}&=&\delta_{\lambda\mu}\mathbf e_\nu,\\
\label{eq:e_munu e_lambda left}
\textsf e_{\mu\nu}\cdot\mathbf e_\lambda&=&\delta_{\nu\lambda}\mathbf e_\mu,
\end{eqnarray}
and
\begin{eqnarray}
\label{eq:e_munu e_mu'nu' right}
\cdot\,\textsf e_{\mu'\nu'}\cdot\textsf e_{\mu\nu}&=&\delta_{\nu'\mu}(\cdot\,\textsf e_{\mu'\nu}),\\
\label{eq:e_mu'nu' e_munu left}
\textsf e_{\mu'\nu'}\cdot\mathbf e_{\mu\nu}\cdot\,&=&\delta_{\nu'\mu}\textsf e_{\mu'\nu}\cdot,
\end{eqnarray}
where $\lambda,\mu,\nu\in\{1,2\}$.  Notice that we can rederive these relations if we adopt the definitions
\begin{eqnarray}
\label{eq:e_munu is e_mu e_nu right}
\cdot\,\textsf e_{\mu\nu}&=&\,\cdot\,\mathbf e_\mu\mathbf e_\nu,\\
\textsf e_{\mu\nu}\cdot\,&=&\mathbf e_\mu\mathbf e_\nu\cdot,
\end{eqnarray}
with the understanding that the dot product takes precedence over the juxtaposition (geometric) product.

Let $\cdot\textsf M$ be a right-acting $2\times 2$ matrix and let $\textsf M^T\cdot$ be its left-acting transpose:
\begin{eqnarray}
\label{eq:M is M_11e_11 + M_12e_12 + M_21e_21 + M_22e_22}
\cdot\,\textsf M&=&M_{11}\textsf e_{11}+M_{12}\textsf e_{12}+M_{21}\textsf e_{21}+M_{22}\textsf e_{22},\\
\label{eq:M transpose is M_11e_11 + M_12e_21 + M_21e_12 + M_22e_22}
\textsf M^T\cdot\,&=&M_{11}\textsf e_{11}+M_{12}\textsf e_{21}+M_{21}\textsf e_{12}+M_{22}\textsf e_{22}.
\end{eqnarray}
The action of these two matrices on the vector 
\begin{equation}
\label{eq:r is x_1e_1 + x_2e_2}
\mathbf r=x_1\mathbf e_1+x_2\mathbf e_2
\end{equation}
are related by
\begin{equation}
\label{eq:r' is M transpose r is r M}
\mathbf r'=\textsf M^T\cdot\mathbf r=\mathbf r\cdot\textsf M,
\end{equation}
where $\mathbf r'$ is another vector.  That is,\cite{SugonFernandezMcNamara_2008_arXiv0811.3680v1_p9}
\begin{eqnarray}
\label{eq:r' is M transpose r is r M matrix}
\begin{pmatrix}
x_1'\\
x_2'
\end{pmatrix}
&=&
\begin{pmatrix}
M_{11}&M_{21}\\
M_{12}&M_{22}
\end{pmatrix}
\begin{pmatrix}
x_1\\
x_2
\end{pmatrix}
\nonumber\\
&=&
\begin{pmatrix}
x_1\\
x_2
\end{pmatrix}
\begin{pmatrix}
M_{11}&M_{12}\\
M_{21}&M_{22}
\end{pmatrix}
\end{eqnarray}
Hence,
\begin{eqnarray}
\label{eq:x_1' is x_1M_11 + x_2M_21}
x_1'&=&x_1M_{11}+x_2M_{21},\\
\label{eq:x_2' is x_2M_21 + x_2M_22}
x_2'&=&x_1M_{12} + x_2M_{22}.
\end{eqnarray}
Notice that the column-column multiplication in the action of right-acting matrices is simpler than the row-column multiplication in that of left-acting matrices.

\section{Matrix Optics}

\subsection{Ray Cliffor}

\begin{figure}[ht]
\begin{center}
\setlength{\unitlength}{1 mm}
\begin{picture}(60,50)(0,-10)
\put(0,0){\line(0,1){15}}
\put(2,6){\small $x_1$}
\put(30,0){\line(0,1){35}}
\put(32,15){\small $x_2$}
\multiput(55,0)(0,5){9}{\line(0,1){2.5}}
\put(57,20){\small $x_3$}
\put(0,0){\line(1,0){30}}
\put(15,-5){\small $s_1$}
\multiput(32.5,0)(5,0){5}{\line(1,0){2.5}}
\put(42,-5){\small $s_2$}
\thicklines
\qbezier(0,15)(30,35)(30,35)
\thinlines
\put(0,15){\line(1,0){10}}
\qbezier(7.000,15.000)(7.000,17.119)(5.824,18.883)
\put(9,17){\small $\alpha_1$}
\thicklines
\qbezier(30,35)(55,45)(55,45)
\thinlines
\qbezier(37.000,35.000)(37.000,36.348)(36.499,37.600)
\put(30,35){\line(1,0){10}}
\put(40,36){\small $\alpha_2$}
\end{picture}
\end{center}
\begin{quote}
\vspace{-0.5cm}
\caption{\footnotesize A paraxial ray with height $x_1$ and inclination angle $\alpha_1$ moves to the right by a distance $s_1$ until it hits an refracting surface.  The height of the ray becomes $x_2$ and its inclination angle changes to $\alpha_2$.}
\label{fig:propagation refraction propagation}
\vspace{-0.5cm}
\end{quote}
\end{figure}
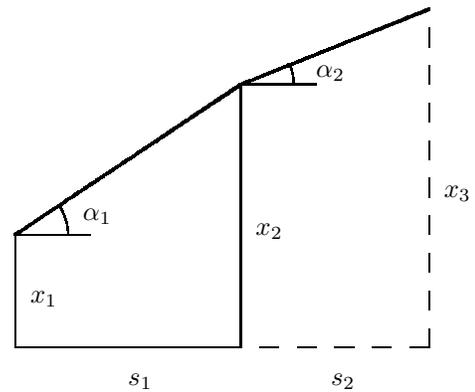

Let us define the complex height-angle vector $\hat r$ as 
\begin{equation}
\label{eq:ray r}
\hat r=x\mathbf e_1+n\alpha i\mathbf e_2=
\begin{pmatrix}
x\\
n\alpha i
\end{pmatrix}
.
\end{equation}
Here, the vector $\mathbf e_3$ as the optical axis pointing to the right, $\mathbf e_1$ as pointing upwards, and $\mathbf e_2$ as pointing out of the paper.  We define the light ray to be moving from left to right.  The height $x$ of the ray is positive if the ray is above the optical axis and negative if below.  The angle $\alpha$ is positive if the ray is inclined and negative if declined.  (Fig.~\ref{fig:propagation refraction propagation})

To define the paraxial angle $\alpha$ more precisely, we use sign functions\cite{SugonMcNamara_2004_ajp72i1pp92-97_p95,SugonMcNamara_2006_AIEP139pp179-224_pp206}.  If $\bm\sigma$ is the direction of propagation of the light ray as it moves close to the direction of the optical axis $\mathbf e_3$, then the ray's angle of inclination $\alpha$ with respect to $\mathbf e_3$ is\cite{SugonMcNamara_2008_arXiv:0810.5224v1_p3}
\begin{equation}
\label{eq:alpha is c_sigmax theta_sigmaz}
\alpha=\theta_{\sigma}\approx c_{\sigma x}\theta_{\sigma z},
\end{equation}
where
\begin{eqnarray}
\label{eq:c_sigma x define}
c_{\sigma x}&=&\frac{\bm\sigma\cdot\mathbf e_1}{|\bm\sigma\cdot\mathbf e_1|},\\
\label{eq:theta_sigma z define}
\theta_{\sigma z}&=&\cos^{-1}(\bm\sigma\cdot\mathbf e_3).
\end{eqnarray}
The sign function $c_{\sigma x}$ is the relative direction of the $\bm\sigma$ along the axis $\mathbf e_1$, with +1 meaning along and $-1$ opposite.  The angle $\theta_{\sigma z}$ is magnitude of the angle between $\bm\sigma$ and $\mathbf e_3$.

\subsection{Matrix Operators}

When light propagates, the paraxial meridional angle $\alpha$ remains constant.  So the ray tracing equations are
\begin{eqnarray}
\label{eq:x' is x + s alpha}
x'&=&x+s\alpha,\\
\label{eq:alpha' is alpha}
\alpha'&=&\alpha,\\
\label{eq:z' is z + c_sigma z s}
z'&=&z+s.
\end{eqnarray}
In most texts, the last equation is assumed.

Using the definition of the ray cliffor $\hat r$ in Eq.~(\ref{eq:ray r}), Eqs.~(\ref{eq:x' is x + s alpha}) and (\ref{eq:alpha' is alpha}) may be combined as
\begin{equation}
\label{eq:r' is r M_S}
\hat r'=\hat r\cdot\textsf M_S=\hat r\cdot(1+\textsf S).
\end{equation}
where
\begin{equation}
\label{eq:S is -iSe_21}
\textsf S=-iS\textsf e_{21}=-i\frac{s}{n}\textsf e_{12}.
\end{equation}
That is,
\begin{equation}
\label{eq:r' is r M_S matrix form}
\begin{pmatrix}
x'\\
n'\alpha' i
\end{pmatrix}
=
\begin{pmatrix}
x\\
n\alpha i
\end{pmatrix}
\begin{pmatrix}
1&0\\
-is/n&1
\end{pmatrix}
.
\end{equation}
Except for the $-i$, the propagation matrix is the same as that given by Hecht\cite{Hecht_1987_Optics_p217}.

On the other hand, when light refracts the height $x$ of the light ray remains constant.  So the ray tracing equations are
\begin{eqnarray}
\label{eq:x' is x}
x'&=&x,\\
\label{eq:n' alpha' is n alpha - Px}
n'\alpha'&=&n\alpha-Px,\\
\label{eq:z' is z}
z'&=&z.
\end{eqnarray}
Here, $P$ is the power of the interface\cite{SugonMcNamara_2008_arXiv:0810.5224v1_p4}:
\begin{equation}
\label{eq:P is c_eta z n - n' over R}
P =c_{\eta z}\frac{n-n'}{R},
\end{equation}
where
\begin{equation}
\label{eq:c_eta z define}
c_{\eta z}=\frac{\bm\eta\cdot\mathbf e_3}{|\bm\eta\cdot\mathbf e_3|}
\end{equation}
is the sign function describing the relative direction of the outward normal vector $\bm\eta$ to the spherical interface of radius $R$, with respect to the optical axis $\mathbf e_3$.  If $c_{\eta z}=+1$, $\bm\eta$ is approximately along $\mathbf e_3$; if $c_{\eta z}=-1$, $\bm\eta$ is opposite.

Using the definition of the ray cliffor $\hat r$ in Eq.~(\ref{eq:ray r}), Eqs.~(\ref{eq:x' is x}) and (\ref{eq:n' alpha' is n alpha - Px}) may be combined into
\begin{equation}
\label{eq:r' is r M_P}
\hat r'=\hat r\cdot\textsf M_P=\hat r\cdot(1+\textsf P),
\end{equation}
where
\begin{equation}
\label{eq:P is -iPe_12}
\textsf P=-iP\textsf e_{12}.
\end{equation}
That is,
\begin{equation}
\label{eq:r' is r M_P matrix form}
\begin{pmatrix}
x'\\
n'\alpha' i
\end{pmatrix}
=
\begin{pmatrix}
x\\
n\alpha i
\end{pmatrix}
\begin{pmatrix}
1&-iP\\
0&1
\end{pmatrix}
.
\end{equation}
Again, except for the $-i$, the refraction matrix is similar to that given by Hecht\cite{Hecht_1987_Optics_p217}.

We may also rewrite the propagation and refraction matrices by using the definition of the exponential of the matrix $\textsf A$ as
\begin{equation}
\label{eq:exp A power series}
e^{\textsf A}=1+\textsf A+\frac{\textsf A^2}{2!}+\frac{\textsf A^3}{3!}+\ldots,
\end{equation}
so that if $\textsf A^2=0$, then
\begin{equation}
\label{eq:exp A is 1 + A}
e^{\textsf A}=1+\textsf A.
\end{equation}
Thus, the exponential of a null-square matrix is the sum of the matrix itself and the unit matrix.

Because $\textsf e_{12}$ and $\textsf e_{21}$ are null-square matrices,
\begin{equation}
\label{eq:e_12 and e_21 squared is 0}
\textsf e_{12}^2=\textsf e_{21}^2=0,
\end{equation}
then we may use Eq.~(\ref{eq:exp A is 1 + A}) to express the matrices $\textsf M_S$ and $\textsf M_P$ in Eqs.~(\ref{eq:r' is r M_S}) and (\ref{eq:r' is r M_P}) as
\begin{eqnarray}
\label{eq:M_S is exp S}
\textsf M_S&=&=e^{\textsf S}=1+\textsf S=1-iS\textsf e_{21},\\
\label{eq:M_P is exp P}
\textsf M_P&=&=e^{\textsf P}=1+\textsf P=1-iP\textsf e_{12}.
\end{eqnarray}
Except for the $-i$ factor, these exponential forms of the propagation and refraction operators were used before by Simon and Wolf\cite{SimonWolf_2000_josav17i2pp342-354_pp345-347}.  These forms are useful when we have a series of propagations or of refractions.  For these cases, we only have to add the arguments of the exponentials, as what we shall see next.

\subsection{System Matrix}

An optical system may be considered a black box: we do not know what is inside.  All we know is that, in general, the relationship between the input ray $\hat r$ and output ray $\hat r'$ is given by
\begin{equation}
\label{eq:r' is r M}
\hat r'=\hat r\cdot\textsf M,
\end{equation}
where $\textsf M$ is a $2\times 2$~matrix.  That is,
\begin{equation}
\label{eq:r' is r M matrix form}
\begin{pmatrix}
x'\\
n'\alpha' i
\end{pmatrix}
=
\begin{pmatrix}
x\\
n\alpha i
\end{pmatrix}
\begin{pmatrix}
A&-iC\\
-iB&D
\end{pmatrix}
,
\end{equation}
so that
\begin{eqnarray}
\label{eq:x' is r M x part}
x'&=&Ax+Bn\alpha,\\
\label{eq:n'alpha' is r M nalpha part}
n'\alpha'&=&-Cx+Dn\alpha.
\end{eqnarray}
Notice that these equations are similar to those in the literature, save for the sign of $C$.

In general, the system matrix $\textsf M$ is a product of propagation and refraction operators (c.f. \cite{Hecht_1987_Optics_p219}):
\begin{equation}
\label{eq:M is product M_Sk M_Pk}
\textsf M=\prod_{k=1}^N \textsf M_{S_k}\textsf M_{P_k}=\prod_{k=1}^N e^{\textsf S_k}e^{\textsf P_k}=\prod_{k=1}^N (1+\textsf S_k)(1+\textsf P_k),
\end{equation}
where
\begin{eqnarray}
\label{eq:matrix S_k}
\textsf S_k&=&-iS_k\textsf e_{21}=-i\frac{s_k}{n_k}\textsf e_{21},\\
\label{eq:matrix P_k}
\textsf P_k&=&-iP_k\textsf e_{12}=-ic_{\eta k}\frac{n_k-n_{k+1}}{R}\,\textsf e_{12}.
\end{eqnarray}
Note that the determinant of $\textsf M$ is unity\cite{BlakerRosenblum_1993_Optics_p37},
\begin{equation}
\label{eq:det M is 1}
|\textsf M|=\prod_{k=1}^N |e^{\textsf S_k}||e^{\textsf P_k}|=1,
\end{equation}
because its factors have a determinant of unity,
\begin{equation}
\label{eq:det of exp is 1}
|e^{\textsf S_k}|=|e^{\textsf P_k}|=1.
\end{equation}

Let us take some special cases.  If $\textsf P_k=0$ for all $k$, then
\begin{equation}
\label{eq:matrix M as product of M_Sk}
\textsf M=\prod_{k=1}^N \textsf M_{Sk}=\prod_{k=1}^N e^{\textsf S_k}=e^{\textsf S}=1+\textsf S,
\end{equation}
where
\begin{equation}
\label{eq:matrix S is sum S_k}
\textsf S=\sum_{k=1}^N \textsf S_k=-i\textsf e_{21}\sum_{k=1}^N S_k.
\end{equation}
In other words, the reduced distance\cite{Meyer-Arendt_1984_IntroductiontoClassicalandModernOptics_p55} or the index-normalized path length $S=s/n$ of a sequence of vertical interfaces (parallel to the $xy-$plane) is equal to the sum of the index-normalized path lengths between successive interfaces.  (Path length is normally defined as $ns$.)

On the other hand, if $\textsf S_k=0$ for all $k$, then
\begin{equation}
\label{eq:matrix M as product of M_Pk}
\textsf M=\prod_{k=1}^N \textsf M_{Pk}=\prod_{k=1}^N e^{\textsf P_k}=e^{\textsf P}=1+\textsf P,
\end{equation}
where
\begin{equation}
\label{eq:matrix P is sum P_k}
\textsf P=\sum_{k=1}^N \textsf P_k=-i\textsf e_{12}\sum_{k=1}^N P_k.
\end{equation}
In other words, the total power $P$ of a sequence of refracting surfaces separated by negligible distances is equal to the sum of the individual powers of each interface.

\subsection{Fermion Combinatorics}

In general, we cannot simply add the arguments of the exponentials in Eq.~(\ref{eq:M is product M_Sk M_Pk}) because 
\begin{equation}
\label{eq:S_k P_k is not equal P_k S_k}
\textsf S_k\textsf P_\ell\neq\textsf P_\ell\textsf S_k.
\end{equation}
for all subscripts $k$ and $\ell$.  So instead of the product-of-exponentials form, we shall use the binomial product form and study how to expedite its expansion.

To expand the binomial product form in Eq.~(\ref{eq:M is product M_Sk M_Pk}), we note several things, which we shall label as rules:  
\medskip

\textbf{Rule 1.}  The only allowed products are those with alternating $\textsf S$ and $\textsf P$ factors, because
\begin{equation}
\label{eq:S_k S_l is P_k P_l is 0}
\textsf S_k\textsf S_\ell=\textsf P_k\textsf P_\ell=0.
\end{equation}

\textbf{Rule 2.}  Since matrix multiplication is not generally commutative, then the order of the factors must be preserved.  This means that the values of the subscript $k$ must be increasing from left to right; if the subscripts are the same, $\textsf S_k$ should come before $\textsf P_k$.

\textbf{Rule 3.}  The matrix $\textsf S_k$ is an $-i\textsf e_{21}$ quantity; $\textsf P_k$ is a $-i\textsf e_{12}$ quantity.  Because the allowed products are those with alternating $\textsf S$ and $\textsf P$ factors, then the Fermion basis of the product depends only on the first and last factors:
\begin{eqnarray}
\label{eq:product start S_k end S_l}
\textsf S_k\cdots\textsf S_\ell,&\quad& (-i)^L\textsf e_{21},\\
\label{eq:product start S_k end P_l}
\textsf S_k\cdots\textsf P_\ell,&\quad& (-i)^L\textsf e_{22},\\
\label{eq:product start P_k end S_l}
\textsf P_k\cdots\textsf S_\ell,&\quad& (-i)^L\textsf e_{11},\\
\label{eq:product start P_k end P_l}
\textsf P_k\cdots\textsf P_\ell,&\quad& (-i)^L\textsf e_{12},
\end{eqnarray} 
where $L$ is the number of factors in the product chain.

\textbf{Rule 4.}  The unit number $1$ is a sum of $\textsf e_{11}$ and $\textsf e_{22}$,
\begin{equation}
\label{eq:unit is e_11 + e_22}
1=\textsf e_{11}+\textsf e_{22}.
\end{equation}
These four rules let us compute the system matrix in a systematic way, by simply listing down the allowed combinations of $\textsf S$ and $\textsf P$.  

\subsection{Thin and Thick Lenses}

To illustrate our four combinatorial rules, let us compute the expansions of $\textsf M$ in Eq.~(\ref{eq:M is product M_Sk M_Pk}) for $k=1$ and $k=2$.
\bigskip

\textbf{Case $k=1$.}  The matrix $\textsf M$ is
\begin{eqnarray}
\label{eq:M for k is 1}
\textsf M&=&(1+\textsf S_1)(1+\textsf P_1)\nonumber\\
&=&1+(\textsf S_1+\textsf P_1)+\,\textsf S_1\textsf P_1.
\end{eqnarray}
In Fermion basis, this is
\begin{eqnarray}
\label{eq:M for k is 1 Fermion basis}
\textsf M&=&\textsf e_{11}(1)+\textsf e_{12}(-i)P_1\nonumber\\
& &+\ \textsf e_{21}(-i)S_1+\textsf e_{22}(1+(-i)^2S_1P_1).
\end{eqnarray}
That is,
\begin{equation}
\label{eq:M for k is 1 matrix}
\textsf M=
\begin{pmatrix}
1&-iP_1\\
-iS_1&1-S_1P_1
\end{pmatrix}
.
\end{equation}
\bigskip

\textbf{Case $k=2$.}  The matrix $\textsf M$ is
\begin{eqnarray}
\label{eq:M for k is 2}
\textsf M&=&(1+\textsf S_1)(1+\textsf P_1)(1+\textsf S_2)(1+\textsf P_2)\nonumber\\
&=&1 +(\textsf S_1+\textsf P_1+\textsf S_2+\textsf P_2)\nonumber\\
& &+\ (\textsf S_1\textsf P_1+\textsf S_1\textsf P_2+\textsf S_2\textsf P_2+\textsf P_1\textsf S_2)\nonumber\\
& &+\ (\textsf S_1\textsf P_1\textsf S_2+\textsf P_1\textsf S_2\textsf P_2)+\textsf S_1\textsf P_1\textsf S_2\textsf P_2.
\end{eqnarray}
In Fermion basis, this is
\begin{eqnarray}
\label{eq:M for k is 2 Fermion basis}
\textsf M&=&\textsf e_{11}(1+(-i)^2P_1S_2)\nonumber\\
& &+\ \textsf e_{12}(-iP_1-iP_2+(-i)^3P_1S_2P_2)\nonumber\\
& &+\ \textsf e_{21}(-iS_1-iS_2+(-i)^3S_1P_1S_2)\nonumber\\
& &+\ \textsf e_{22}(1+(-i)^2S_1P_1+(-i)^2S_1P_2+(-i)^2S_2P_2\nonumber\\
& &\qquad\quad+\ (-i)^4S_1P_1S_2P_2).
\end{eqnarray}
That is,
\begin{equation}
\label{eq:M for k is 2 matrix}
\textsf M=
\begin{pmatrix}
(1-P_1S_2)&-i(P_1+P_2-P_1S_2P_2)\\
\\
-i
\begin{pmatrix}
S_1+S_2\\
-S_1P_1S_2
\end{pmatrix}
&
\begin{pmatrix}
1-S_1P_1-S_1P_2-S_2P_2\\
+\ S_1P_1S_2P_2
\end{pmatrix}
\end{pmatrix}
\end{equation}

\textbf{Check.}  To check our computations, we set $S_1=0$ in Eq.~(\ref{eq:M for k is 2 matrix}) to get
\begin{equation}
\label{eq:M_thick}
\textsf M_{thick}=
\begin{pmatrix}
(1-P_1S_2)&-i(P_1+P_2-P_1S_2P_2)\\
\\
-iS_2&(1-S_2P_2)
\end{pmatrix}
,
\end{equation}
Notice that except for the $-i$, Eq.~(\ref{eq:M_thick}) is similar in form to the system matrix for a thick lens as given in Klein and Furtak (they use left-acting matrices and angle-height vectors)\cite{KleinFurtak_1986_Optics_p156}.  

Furthermore, if we set $S_2=0$ in Eq.~(\ref{eq:M_thick}), then we obtain the corresponding thin lens matrix\cite{KleinFurtak_1986_Optics_p157}:
\begin{equation}
\label{eq:M_thin}
\textsf M_{thin}=
\begin{pmatrix}
1&-i(P_1+P_2)\\
0&1
\end{pmatrix}
\end{equation}
Comparison of Eq.~(\ref{eq:M_thin}) with the refraction matrix operator $\textsf M_P$ in Eq.~(\ref{eq:r' is r M_P matrix form}) yields the thin lens power 
\begin{equation}
\label{eq:P is P_1+P_2}
P=P_1+P_2,
\end{equation}
which is what we expect.  That is, the total power of a thin lens is equal to the sum of the powers of its refracting surfaces.

\section{Optical Systems}

So far, we have considered the input and output rays to be very close to the optical black box.  We shall now relax this restriction.

Let the matrix $\textsf M_{\textrm{box}}$ describe a black box:
\begin{equation}
\label{eq:M_box}
\textsf M_{\textrm{box}}=
\begin{pmatrix}
M_{11}&-iM_{22}\\
-iM_{21}&M_{22}
\end{pmatrix}
,
\end{equation}
which satisfies Eq.~(\ref{eq:det M is 1}),
\begin{equation}
\label{eq:det M_box is 1}
|\textsf M_{\textrm{box}}|=M_{11}M_{22}+M_{12}M_{21}=1.
\end{equation}
Notice that unlike the determinant of standard system matrices with real coefficients, the determinant of $\textsf M_{\textrm{box}}$ in Eq.~(\ref{eq:det M_box is 1}) is not a difference but a sum.

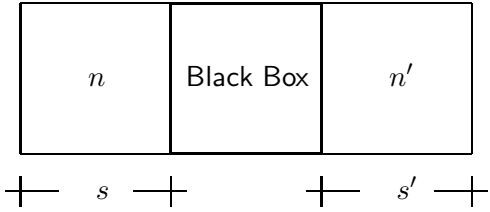
\begin{figure}[h]
\begin{center}
\setlength{\unitlength}{1 mm}
\begin{picture}(70,35)(-25,-10)
\thinlines
\multiput(-20,0)(0,20){2}{\line(1,0){60}}
\multiput(-20,0)(60,0){2}{\line(0,1){20}}
\thicklines
\multiput(0,0)(20,0){2}{\line(0,1){20}}
\multiput(0,0)(0,20){2}{\line(1,0){20}}
\put(2,9){\textsf{Black Box}}
\put(-11,9){$n$}
\put(29,9){$n'$}
\thinlines
\multiput(-20,-7)(20,0){4}{\line(0,1){4}}
\multiput(-22,-5)(17,0){2}{\line(1,0){7}}
\put(-10,-6){$s$}
\multiput(18,-5)(17,0){2}{\line(1,0){7}}
\put(30,-6){$s'$}
\end{picture}
\end{center}
\begin{quote}
\vspace{-0.5cm}
\caption{\footnotesize An optical system with a black box.  The input side is at the distance $s$ to the left of the box; the output side is at a distance of $s'$ to the right of the box.}
\label{fig:black box}
\vspace{-0.5cm}
\end{quote}
\end{figure}

To the left of the box at the reduced distance $S=s/n$ is the input ray characterized by height $x$ and optical angle $n\alpha$.  To the right of the box at an reduced distance $S'=s'/n'$ is the output ray characterized by height $x'$ and dilated angle $n'\alpha'$.  In other words, the system matrix $\textsf M$ in Eq.~(\ref{eq:r' is r M matrix form}) is 
\begin{eqnarray}
\label{eq:M is M_S M_box M_S'}
\textsf M\!\!&=&\!\!\!\!
\begin{pmatrix}
A&-iC\\
-iB&D
\end{pmatrix}
=\textsf M_{S}\textsf M_{\textrm{box}}\textsf M_{S^\prime}\nonumber\\
&=&\!\!\!\!
\begin{pmatrix}
1&0\\
-iS&1
\end{pmatrix}
\begin{pmatrix}
M_{11}&-iM_{12}\\
-iM_{21}&M_{22}
\end{pmatrix}
\begin{pmatrix}
1&0\\
-iS'&1
\end{pmatrix}
.
\end{eqnarray}
Except for the $-i$, this matrix product is similar in form to that of Simmons and Guttmann\cite{SimmonsGuttmann_1970_StatesWavesandPhotons_p8}.  (In Ghatak\cite{Ghatak_1977_Optics_p52} and in Nussbaum and Phillips\cite{NussbaumPhillips_1976_ContemporaryOpticsforScientistsandEngineers_p52} the matrix coefficients of $\textsf M_{\textrm{box}}$ are labeled by $-a$, $b$, $c$, and $-d$.)

From Eq.~(\ref{eq:M is M_S M_box M_S'}) we can arrive at two conclusions.  First, because the sytem matrix is a product of matrices with unit determinants, then 
\begin{equation}
\label{eq:det M_S is AB + CD is 1}
|\textsf{M}_S|=AB+CD=1.
\end{equation}
Second, the elements of the system matrix are given by 
\begin{eqnarray}
\label{eq:A element}
A&=&M_{11}-M_{12}S',\\
\label{eq:B element}
B&=&M_{21}+M_{22}S' +M_{11}S-M_{12}SS',\\
\label{eq:C element}
C&=&M_{12},\\
\label{eq:D element}
D&=&M_{22}-M_{12}S.
\end{eqnarray}

Now, the input and output rays are related to the system matrix $\textsf M$ by Eqs.~(\ref{eq:x' is r M x part}) and (\ref{eq:n'alpha' is r M nalpha part}).  Our aim is to use these equations to give a geometric interpretation to the $\textsf M_{\textrm{box}}$ parameters $M_{11}$, $M_{12}$, $M_{21}$, and $M_{22}$.  To do this, we shall choose from the set $\{x,x^\prime,\alpha,\alpha^\prime\}$ a pair of variables and set the other variables constant.  We shall consider four cases:
\begin{eqnarray}
\label{eq:x' is x' of x}
x^\prime&=x'(x);\quad&\alpha,\alpha'=\textrm{constants}\\
\label{eq:x' is x' of alpha}
x^\prime&=x'(\alpha);\quad&x,\alpha'=\textrm{constants}\\
\label{eq:alpha' is alpha' of x}
\alpha'&=\alpha'(x);\quad&\alpha,x'=\textrm{constants}\\
\label{eq:alpha' is alpha' of alpha}
\alpha'&=\alpha'(\alpha);\quad&x,x^\prime=\textrm{constants}.
\end{eqnarray}

\subsection{Telescopic System}

\begin{figure}[ht]
\begin{center}
\setlength{\unitlength}{1 mm}
\begin{picture}(70,35)(-25,-5)
\thicklines
\multiput(0,0)(0,20){2}{\line(1,0){20}}
\multiput(0,0)(20,0){2}{\line(0,1){20}}
\put(2,9){\textsf{Black Box}}
\thinlines
\qbezier(-20,5)(0,10)(0,10)
\qbezier(-2,8)(0,10)(0,10)
\qbezier(-2,11)(0,10)(0,10)
\qbezier(-20,10)(0,15)(0,15)
\qbezier(-2,13)(0,15)(0,15)
\qbezier(-2,16)(0,15)(0,15)
\multiput(-27,10)(0,-5){2}{\line(1,0){4}}
\put(-26,6){\small{$\delta x$}}
\multiput(-25,15)(0,-10){2}{\line(0,-1){5}}
\qbezier(-26,11)(-25,10)(-25,10)
\qbezier(-24,11)(-25,10)(-25,10)
\qbezier(-26,4)(-25,5)(-25,5)
\qbezier(-24,4)(-25,5)(-25,5)
\put(-20,5){\line(1,0){8}}
\qbezier(-15.000,5.000)(-15.000,5.616)(-15.149,6.213)
\put(-10,4){\small{$\alpha$}}
\thinlines
\qbezier(20,5)(40,10)(40,10)
\qbezier(38,8)(40,10)(40,10)
\qbezier(38,11)(40,10)(40,10)
\qbezier(20,10)(40,15)(40,15)
\qbezier(38,13)(40,15)(40,15)
\qbezier(38,16)(40,15)(40,15)
\multiput(43,15)(0,-5){2}{\line(1,0){4}}
\put(44,11){\small{$\delta x'$}}
\multiput(45,20)(0,-10){2}{\line(0,-1){5}}
\qbezier(44,9)(45,10)(45,10)
\qbezier(46,9)(45,10)(45,10)
\qbezier(44,16)(45,15)(45,15)
\qbezier(46,16)(45,15)(45,15)
\put(20,5){\line(1,0){8}}
\qbezier(25.000,5.000)(25.000,5.616)(24.851,6.213)
\put(30,4){\small{$\alpha'$}}
\end{picture}
\end{center}
\begin{quote}
\vspace{-0.5cm}
\caption{\footnotesize In a telescopic system, parallel rays emerge as parallel rays.}
\label{fig:telescopic system}
\vspace{-0.5cm}
\end{quote}
\end{figure}
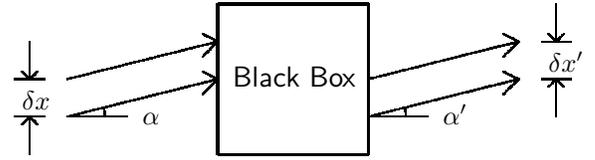

If $x'=x'(x)$, with the the angles $\alpha$ and $\alpha'$ constants, then the input and output beams are both bundles of parallel rays---a telescopic or afocal system.  If this is true, then we may differentiate Eqs.~(\ref{eq:x' is r M x part}) and (\ref{eq:n'alpha' is r M nalpha part}) with respect to $x$ to obtain\cite{Goodman_1995_HandbookofOpticsI_ch1p71,NazarathyShamir_1982_josav72i3pp356-364_p362}
\begin{eqnarray}
\label{eq:d x' dx is A}
\frac{\partial x'}{\partial x}&=A&=M_{11}-M_{12}S',\\
\label{eq:C is 0}
0&=C&=M_{12}.
\end{eqnarray}
Using these equations, together with Eqs.~(\ref{eq:x' is r M x part}) to (\ref{eq:n'alpha' is r M nalpha part}) and (\ref{eq:A element}) to (\ref{eq:D element}), we arrive at
\begin{eqnarray}
\label{eq:d x' d x is M_11}
\frac{\partial x'}{\partial x}&=&M_{11},\\
\label{eq:n alpha' over n alpha is M_12}
\frac{n'\alpha'}{n\alpha}&=&M_{22}.
\end{eqnarray}
Thus, in a telescopic system, the ratio of the outgoing and ingoing rays's index-dilated angles is the angular magnification $M_{22}$\cite{SimmonsGuttmann_1970_StatesWavesandPhotons_p11}; the ratio of the change in the output and input heights is $M_{11}$.  

If we multiply Eqs.~(\ref{eq:d x' d x is M_11}) and (\ref{eq:n alpha' over n alpha is M_12}), we get
\begin{equation}
\label{eq:Lagrange theorem telescopic}
\frac{n'\alpha'}{n\alpha}\frac{\partial x'}{\partial x}=1=|\textsf M_{\textrm box}|,
\end{equation}
because $M_{12}=0$.  Equation~(\ref{eq:Lagrange theorem telescopic}) is the Lagrange's theorem for a telescopic system.  An alternative formulation of this theorem is
\begin{equation}
\label{eq:n' alpha' delta x' is n alpha delta x}
n'\alpha'\,\delta x' = n\alpha\,\delta x,
\end{equation}
where $n\alpha\,\delta x$ is the corresponding Lagrange invariant.  That is, if the angles of the input and output rays are positive constants, then the change in height $x$ of the input ray is proportional to the change in height of the output ray $x'$.

\subsection{Inverse Fourier Transforming\\ 
System}

\begin{figure}[ht]
\begin{center}
\setlength{\unitlength}{1 mm}
\begin{picture}(70,35)(-25,-5)
\thicklines
\multiput(0,0)(0,20){2}{\line(1,0){20}}
\multiput(0,0)(20,0){2}{\line(0,1){20}}
\put(2,9){\textsf{Black Box}}
\thinlines
\qbezier(-25,15)(0,17.5)(0,17.5)
\qbezier(-2,18.5)(0,17.5)(0,17.5)
\qbezier(-2,16)(0,17.5)(0,17.5)
\qbezier(-25,15)(0,5)(0,5)
\qbezier(-2,4.5)(0,5)(0,5)
\qbezier(-1,7)(0,5)(0,5)
\qbezier(-20.025,15.498)(-19.903,14.280)(-20.358,13.143)
\put(-16,12){\small{$\delta\alpha$}}
\multiput(-25,5)(0,10){2}{\line(0,1){5}}
\qbezier(-26,16)(-25,15)(-25,15)
\qbezier(-24,16)(-25,15)(-25,15)
\multiput(-27,10)(0,5){2}{\line(1,0){4}}
\qbezier(-26,9)(-25,10)(-25,10)
\qbezier(-24,9)(-25,10)(-25,10)
\put(-26,11.5){\small{$x$}}
\thinlines
\qbezier(20,5)(40,10)(40,10)
\qbezier(38,8)(40,10)(40,10)
\qbezier(38,11)(40,10)(40,10)
\qbezier(20,10)(40,15)(40,15)
\qbezier(38,13)(40,15)(40,15)
\qbezier(38,16)(40,15)(40,15)
\multiput(43,15)(0,-5){2}{\line(1,0){4}}
\put(44,11){\small{$\delta x'$}}
\multiput(45,20)(0,-10){2}{\line(0,-1){5}}
\qbezier(44,9)(45,10)(45,10)
\qbezier(46,9)(45,10)(45,10)
\qbezier(44,16)(45,15)(45,15)
\qbezier(46,16)(45,15)(45,15)
\put(20,5){\line(1,0){8}}
\qbezier(25.000,5.000)(25.000,5.616)(24.851,6.213)
\put(30,4){\small{$\alpha'$}}
\end{picture}
\end{center}
\begin{quote}
\vspace{-0.5cm}
\caption{\footnotesize In an inverse Fourier transforming system, rays moving out from a point source become parallel rays.}
\label{fig:inverse Fourier transforming system}
\vspace{-0.5cm}
\end{quote}
\end{figure}
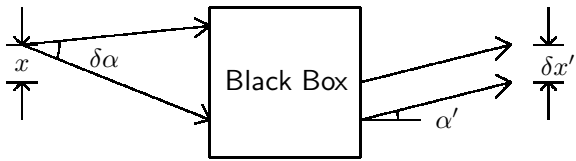

If $x'=x'(\alpha)$, with $x$ and $\alpha$ constants, then the system transforms a point source into a beam of paralell rays---an inverse Fourier transforming system.  If this is true, then we may differentiate Eqs.~(\ref{eq:x' is r M x part}) and (\ref{eq:n'alpha' is r M nalpha part}) with respect to the index-dilated angle $n\alpha$ to obtain\cite{Goodman_1995_HandbookofOpticsI_ch1p71,NazarathyShamir_1982_josav72i3pp356-364_p362}
\begin{eqnarray}
\label{eq:d x' d n alpha is B}
\frac{\partial x'}{\partial (n\alpha)}&=B=&M_{21}+M_{22}S'\nonumber\\
& &+\ M_{11}S-M_{12}SS',\\
\label{eq:D is 0}
0&=D=&M_{22}-M_{12}S.
\end{eqnarray}
Using these equations, together with Eqs.~(\ref{eq:x' is r M x part}) to (\ref{eq:n'alpha' is r M nalpha part}) and (\ref{eq:A element}) to (\ref{eq:D element}), we arrive at
\begin{eqnarray}
\label{eq:S is M_22 over M_{12}}
S&=&\frac{M_{22}}{M_{12}}\equiv f,\\
\label{eq:n alpha' over x is -M_12}
\frac{n'\alpha'}{x}&=&-M_{12},\\
\label{eq:d x' d n alpha is det M_box over M_12}
\frac{\partial x'}{\partial(n\alpha)}&=&\frac{|\textsf M_{\textrm{box}}|}{M_{12}}=\frac{1}{M_{12}}.
\end{eqnarray}
Thus, in an inverse Fourier transforming system, a point light source at the input focal distance $f=M_{22}/M_{12}$ to the left of the black box transforms to a bundle of parallel rays at the output\cite{SimmonsGuttmann_1970_StatesWavesandPhotons_p10}; the ratio of the optical inclination angle $n'\alpha'$ of the output beam to the height $x$ of the point source is $-M_{12}$; the change in the height $x'$ of the output ray with respect to the change in the dilated angle $n\alpha$ of the input ray is $1/M_{12}$.

If we multiply Eqs.~(\ref{eq:n alpha' over x is -M_12}) and (\ref{eq:d x' d n alpha is det M_box over M_12}), we get
\begin{equation}
\label{eq:Lagrange theorem inverse Fourier}
\frac{n'\alpha'}{x}\frac{\partial x'}{\partial(n\alpha)}=-1.
\end{equation}
Equation~(\ref{eq:Lagrange theorem inverse Fourier}) is the Lagrange's theorem for an inverse Fourier transforming system.  In terms of differentials, Eq.~(\ref{eq:Lagrange theorem inverse Fourier}) may be written as
\begin{equation}
\label{eq:n' alpha' delta x' is -x delta n alpha}
n'\alpha'\,\delta x'=-x\, \delta (n\alpha).
\end{equation}
Notice that though we could not define a corresponding Lagrange invariant for this system, we can still provide a geometrical interpretation to Eq.~(\ref{eq:n' alpha' delta x' is -x delta n alpha}): if the input height $x$ and the output angle $\alpha'$ are positive constants, then a positive change in the angle $\alpha$ of the input ray would result to a negative change in the output height $x'$ of the output ray.

\subsection{Fourier Transforming System}

\begin{figure}[ht]
\begin{center}
\setlength{\unitlength}{1 mm}
\begin{picture}(70,35)(-25,-5)
\thicklines
\multiput(0,0)(0,20){2}{\line(1,0){20}}
\multiput(0,0)(20,0){2}{\line(0,1){20}}
\put(2,9){\textsf{Black Box}}
\thinlines
\qbezier(-20,5)(0,10)(0,10)
\qbezier(-2,8)(0,10)(0,10)
\qbezier(-2,11)(0,10)(0,10)
\qbezier(-20,10)(0,15)(0,15)
\qbezier(-2,13)(0,15)(0,15)
\qbezier(-2,16)(0,15)(0,15)
\multiput(-27,10)(0,-5){2}{\line(1,0){4}}
\put(-26,6){\small{$\delta x$}}
\multiput(-25,15)(0,-10){2}{\line(0,-1){5}}
\qbezier(-26,11)(-25,10)(-25,10)
\qbezier(-24,11)(-25,10)(-25,10)
\qbezier(-26,4)(-25,5)(-25,5)
\qbezier(-24,4)(-25,5)(-25,5)
\put(-20,5){\line(1,0){8}}
\qbezier(-15.000,5.000)(-15.000,5.616)(-15.149,6.213)
\put(-10,4){\small{$\alpha$}}
\thinlines
\qbezier(20,2.5)(45,5)(45,5)
\qbezier(28,4.5)(30,3.5)(30,3.5)
\qbezier(28,2)(30,3.5)(30,3.5)
\qbezier(20,15)(45,5)(45,5)
\qbezier(29,13)(30,11)(30,11)
\qbezier(28,10)(30,11)(30,11)
\qbezier(40.025,4.502)(39.903,5.720)(40.358,6.857)
\put(32,5){\small{$\delta\alpha'$}}
\multiput(45,0)(0,10){2}{\line(0,1){5}}
\multiput(43,5)(0,5){2}{\line(1,0){4}}
\qbezier(44,4)(45,5)(45,5)
\qbezier(46,4)(45,5)(45,5)
\qbezier(44,11)(45,10)(45,10)
\qbezier(46,11)(45,10)(45,10)
\put(45,6.5){\small{$x'$}}

\end{picture}
\end{center}
\begin{quote}
\vspace{-0.5cm}
\caption{\footnotesize In a Fourier transforming system, parallel rays converge to a point.  Note that the actual angle $\delta\alpha'$ is the angle opposite to the one labeled in the figure.}
\label{fig:Fourier transforming system}
\vspace{-0.5cm}
\end{quote}
\end{figure}
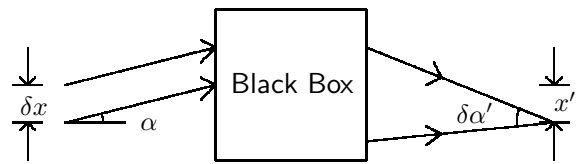

If $\alpha'=\alpha'(x)$, with $x'$ and $\alpha$ constants, then the system focuses a beam of parallel rays into a point---a Fourier transforming system.  This means that we may differentiate Eqs.~(\ref{eq:x' is r M x part}) and (\ref{eq:n'alpha' is r M nalpha part}) with respect to $x$ to obtain\cite{Goodman_1995_HandbookofOpticsI_ch1p71,NazarathyShamir_1982_josav72i3pp356-364_p362}
\begin{eqnarray}
\label{eq:A is 0}
0&=A&=M_{11}-M_{12}S',\\
\label{eq:d n alpha' d x is -M_12}
\frac{\partial(n'\alpha')}{\partial x}&=-C&=-M_{12}.
\end{eqnarray}
Using these equations, together with Eqs.~(\ref{eq:x' is r M x part}) to (\ref{eq:n'alpha' is r M nalpha part}) and (\ref{eq:A element}) to (\ref{eq:D element}), we arrive at
\begin{eqnarray}
\label{eq:S' is M_11 over M_12}
S'&=&\frac{M_{11}}{M_{12}}\equiv f',\\
\label{eq:x' over n alpha is M_box over M_12}
\frac{x'}{n\alpha}&=&\frac{|\textsf M_{\textrm{box}}|}{M_{12}}=\frac{1}{M_{12}}.
\end{eqnarray}
Notice that Eqs.~(\ref{eq:S' is M_11 over M_12}), (\ref{eq:x' over n alpha is M_box over M_12}), and (\ref{eq:d n alpha' d x is -M_12}) are the conjugate relations for Eqs.~(\ref{eq:S is M_22 over M_{12}}) to (\ref{eq:d x' d n alpha is det M_box over M_12}).  That is, in a Fourier transforming system, a bundle of parallel rays at the input is focused to a point source at the output at the focal distance $f'=M_{11}/M_{12}$ to the right of the black box; the ratio of the height $x$ of the focus to the the dilated angle $n\alpha$ of the input beam is $-M_{12}$; the change in the dilated angle $n'\alpha'$ at the output focus with respect to the change in the height $x$ of the input beam is $1/M_{12}$.

If we multiply Eqs.~(\ref{eq:d n alpha' d x is -M_12}) and (\ref{eq:x' over n alpha is M_box over M_12}), we get
\begin{equation}
\label{eq:Lagrange theorem Fourier}
\frac{x'}{n\alpha}\frac{\partial(n'\alpha')}{\partial x}=-1.
\end{equation}
Equation~(\ref{eq:Lagrange theorem Fourier}) is the Lagrange's theorem for a Fourier transforming system.  In terms of differentials, Eq.~(\ref{eq:Lagrange theorem Fourier}) may be written as
\begin{equation}
\label{eq:x' delta n' alpha' is -n alpha delta x}
x'\,\delta(n'\alpha')=-n\alpha\,\delta x.
\end{equation}
Notice that though we could not also define a corresponding Lagrange invariant for this system, we could also still interpret Eq.~(\ref{eq:x' delta n' alpha' is -n alpha delta x}) geometrically: if the input ray angle $\alpha$ and the output height $x'$ are positive constants, then a positive change in the input ray height $x$ would result to a negative change in the output ray angle $\alpha'$.

\subsection{Imaging System}

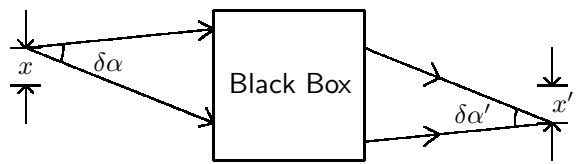
\begin{figure}[hb]
\begin{center}
\setlength{\unitlength}{1 mm}
\begin{picture}(70,35)(-25,-5)
\thicklines
\multiput(0,0)(0,20){2}{\line(1,0){20}}
\multiput(0,0)(20,0){2}{\line(0,1){20}}
\put(2,9){\textsf{Black Box}}
\thinlines
\qbezier(-25,15)(0,17.5)(0,17.5)
\qbezier(-2,18.5)(0,17.5)(0,17.5)
\qbezier(-2,16)(0,17.5)(0,17.5)
\qbezier(-25,15)(0,5)(0,5)
\qbezier(-2,4.5)(0,5)(0,5)
\qbezier(-1,7)(0,5)(0,5)
\qbezier(-20.025,15.498)(-19.903,14.280)(-20.358,13.143)
\put(-16,12){\small{$\delta\alpha$}}
\multiput(-25,5)(0,10){2}{\line(0,1){5}}
\qbezier(-26,16)(-25,15)(-25,15)
\qbezier(-24,16)(-25,15)(-25,15)
\multiput(-27,10)(0,5){2}{\line(1,0){4}}
\qbezier(-26,9)(-25,10)(-25,10)
\qbezier(-24,9)(-25,10)(-25,10)
\put(-26,11.5){\small{$x$}}
\thinlines
\qbezier(20,2.5)(45,5)(45,5)
\qbezier(28,4.5)(30,3.5)(30,3.5)
\qbezier(28,2)(30,3.5)(30,3.5)
\qbezier(20,15)(45,5)(45,5)
\qbezier(29,13)(30,11)(30,11)
\qbezier(28,10)(30,11)(30,11)
\qbezier(40.025,4.502)(39.903,5.720)(40.358,6.857)
\put(32,5){\small{$\delta\alpha'$}}
\multiput(45,0)(0,10){2}{\line(0,1){5}}
\multiput(43,5)(0,5){2}{\line(1,0){4}}
\qbezier(44,4)(45,5)(45,5)
\qbezier(46,4)(45,5)(45,5)
\qbezier(44,11)(45,10)(45,10)
\qbezier(46,11)(45,10)(45,10)
\put(45,6.5){\small{$x'$}}

\end{picture}
\end{center}
\begin{quote}
\vspace{-0.5cm}
\caption{\footnotesize In an imaging system, rays leaving a point source converge to a point.  Note that the actual angle $\delta\alpha'$ is the angle opposite to the one labeled in the figure.}
\label{fig:imaging system}
\vspace{-0.5cm}
\end{quote}
\end{figure}

If $\alpha'=\alpha'(\alpha)$, with $x$ and $x'$ constants, then the system transforms light from a point source to another point source---an imaging system.  If this is true, then we may differentiate Eqs.~(\ref{eq:x' is r M x part}) and (\ref{eq:n'alpha' is r M nalpha part}) with respect to the input optical angle $n\alpha$ to obtain\cite{Goodman_1995_HandbookofOpticsI_ch1p71,NazarathyShamir_1982_josav72i3pp356-364_p362}
\begin{eqnarray}
\label{eq:B is 0 is M_21 + M_22 S'}
0&=B=&M_{21}+M_{22}S'\nonumber\\
& &+\ M_{11}S-M_{12}SS',\\
\label{eq:d n alpha' d n alpha is D}
\frac{\partial(n'\alpha')}{\partial(n\alpha)}&=D=&M_{22}-M_{12}S.
\end{eqnarray}
Using these equations, together with Eqs.~(\ref{eq:x' is r M x part}) to (\ref{eq:n'alpha' is r M nalpha part}) and (\ref{eq:A element}) to (\ref{eq:D element}), we arrive at\cite{NussbaumPhillips_1976_ContemporaryOpticsforScientistsandEngineers_p16,Ghatak_1977_Optics_p56}
\begin{eqnarray}
\label{eq:S' is Moebius S}
S'&=&\frac{M_{11}S+M_{21}}{M_{12}S-M_{22}},\\
\label{eq:S is Moebius S'}
S&=&\frac{M_{22}S'+M_{21}}{M_{12}S'-M_{11}},
\end{eqnarray}
and 
\begin{equation}
\label{eq:x' over x is magnification}
\xi=\frac{x'}{x}=M_{11}-M_{12}S'=\frac{-1}{M_{12}S-M_{22}},
\end{equation}
where we used the identity $|\textsf M_{\textrm box}|=1$.  Equation~(\ref{eq:S' is Moebius S}) are the Moebius relations for the object and image distances.  Equation (\ref{eq:x' over x is magnification}) is the definition for lateral magnification\cite{SimmonsGuttmann_1970_StatesWavesandPhotons_p12}.

There are two relations that we can derive from these equations:

First, the product of Eqs.~(\ref{eq:d n alpha' d n alpha is D}) and (\ref{eq:x' over x is magnification}) is
\begin{equation}
\label{eq:Lagrange theorem imaging}
\frac{x'}{x}\frac{\partial(n'\alpha')}{\partial(n\alpha)}=-1,
\end{equation}
which is the Lagrange's theorem for the imaging system.  In terms of differentials, Eq.~(\ref{eq:Lagrange theorem imaging}) may be written as
\begin{equation}
\label{eq:x' delta n' alpha' is -x delta n alpha}
x'\,\delta(n'\alpha')=-x\,\delta(n\alpha)
\end{equation}
Notice that because of the presence of the negative sign, the quantity $x\delta(n\alpha)$ is not a Lagrange invariant; rather, it is its magnitude $|x\delta(n\alpha)|$.  The geometrical interpretation of Eq.~(\ref{eq:x' delta n' alpha' is -x delta n alpha}) is as follows: if the input and output ray heights are positive constants, then a positive change in the input angle $\alpha$ would result to a negative change in the output angle $\alpha'$.  (Note: the paraxial angle $\alpha$ is measured from the optical axis $\mathbf e_3$ pointing to the right.  A positive $\alpha$ is a counterclockwise rotation; a negative $\alpha$ is clockwise.  Thus, the actual $\delta\alpha'$ in Fig.~\ref{fig:imaging system} is the angle opposite to the one labeled in the figure, i.e., to the right of the image focus.)

Second, we may rewrite Eq.~(\ref{eq:x' over x is magnification}) as
\begin{equation}
\label{eq:product MS - M is M_box}
(M_{12}S-M_{22})(M_{12}S'-M_{11})=|\textsf M_{\textrm{box}}|=1.
\end{equation}
In terms of the focal lengths $f$ in Eq.~(\ref{eq:S is M_22 over M_{12}}) and $f'$ in Eq.~(\ref{eq:S' is M_11 over M_12}), Eq.~(\ref{eq:product MS - M is M_box}) becomes
\begin{equation}
\label{eq:Z_S Z_S' is 1}
(S-f)(S'-f')=ZZ'=\frac{1}{M_{12}^2}.
\end{equation}
This is similar to the Newton's lens equation\cite{Meyer-Arendt_1984_IntroductiontoClassicalandModernOptics_p48}
\begin{equation}
\label{eq:ZZ'=FF'}
ZZ'=FF'.
\end{equation}
Note that the focal lengths $f$ and $f'$ are measured from the left-most and right-most side of the optical black box, while $F$ and $F'$ are measured from the left and right gaussian planes.

\subsection{Summary}
The system matrix $\textsf M$ is defined as the ABCD matrix operating on input ray vector $\hat r$ to yield the output ray vector $\hat r'$
\begin{equation}
\label{eq:r' is r M summary}
\begin{pmatrix}
x'\\
in'\alpha'
\end{pmatrix}
=
\begin{pmatrix}
x\\
in\alpha
\end{pmatrix}
\begin{pmatrix}
A&-iC\\
-iB&D
\end{pmatrix}
,
\end{equation}
where
\begin{eqnarray}
\label{A C partial summary}
A=\frac{\partial x'}{\partial x},\quad\,&\quad& C=-\frac{\partial(n'\alpha')}{\partial x},\\
\label{B D partial summary}
B=\frac{\partial x'}{\partial (n\alpha)},&\quad& D=\frac{\partial(n'\alpha')}{\partial(n\alpha)}.
\end{eqnarray}
Because the system matrix has a unit determinant, then
\begin{equation}
\label{eq:1 is AD+CB partial summary}
1=\frac{\partial x'}{\partial x}\frac{\partial(n'\alpha')}{\partial(n\alpha)}-\frac{\partial(n'\alpha')}{\partial x}\frac{\partial x'}{\partial (n\alpha)},
\end{equation}
as given by Goodman.\cite{Goodman_1995_HandbookofOpticsI_ch1p45}

The four Lagrange theorems for the four optical system types may be combined into one:
\begin{eqnarray}
\label{eq:Lagrange theorem summary}
1&=&\frac{n'\alpha'}{n\alpha}\frac{\partial x'}{\partial x};\quad\qquad\alpha,\alpha'=const.\nonumber\\
&=&-\frac{n'\alpha'}{x}\frac{\partial x'}{\partial(n\alpha)};\quad x,\alpha'=const.\nonumber\\
&=&-\frac{x'}{n\alpha}\frac{\partial(n'\alpha')}{\partial x};\quad \alpha,x'=const.\nonumber\\
&=&-\frac{x'}{x}\frac{\partial(n'\alpha')}{\partial(n\alpha)};\ \quad x,x'=const.
\end{eqnarray}
We shall refer to Eq.~(\ref{eq:Lagrange theorem summary}) as the unified Lagrange theorem for optical systems.

\section{Conclusions}
We used the orthonormality axiom in geometric algebra to show that the height-angle vector of a light ray must be complex.  We proposed that the matrix operators to this vector should be right-acting, in order to follow the sequence of surfaces traversed by a light ray as it moves from left-to-right close to the optical axis.  We showed that the propagation and refraction matrix operators may be expressed as a sum of a unit matrix and a imaginary non-diagonal Fermion matrix.  We developed combinatorial rules for finding the product of a succession of propagation and refraction matrices, without doing explicit matrix multiplication.  This product is the right-acting system matrix with real and imaginary coefficients.

We factored out the system matrix as a product of the input propagation matrix, the black box matrix, and the output propagation matrix.  Based on the coefficients of the system matrix, we classified the optical systems into four: telescopic, inverse Fourier transforming, Fourier transforming, and imaging.  We showed that all four systems have a corresponding Lagrange theorem expressed in partial derivatives, and that these theorems may be combined into one.  We transformed these theorems in terms of Lagrange differentials, which allows us to geometrically interpret the effect of the change in input variable to the change in the output variable.  We showed that these differential relations result to the Lagrange invariants only for the telescopic and imaging systems.

In a future work, we shall revisit paraxial skew ray tracing using complex vectors and right-acting matrices.  

\section*{\small{Acknowledgments}}
This research was supported by the Manila Observatory and by the Physics Department of Ateneo de Manila University.


\begin{thebibliography}{99}
\footnotesize
\bibitem{NussbaumPhillips_1976_ContemporaryOpticsforScientistsandEngineers_p10}
Allen Nussbaum and Richard A. Phillips, \textsl{Contemporary Optics for Scientists and Engineers} (Prentice-Hall, Englewood Cliffs, NJ, 1976), p. 10.

\bibitem{BaylisHuschiltWei_1992_ajp60i9pp788-797_p789}
William E. Baylis, J. Huschilt, and Jiansu Wei, ``Why i?'' Am. J. Phys. \textbf{60}(9), 788--797 (1992).  See p. 789.  Baylis et al's form is $\mathbf e_j\mathbf e_k+\mathbf e_k\mathbf e_j=2\delta_{jk}$.

\bibitem{Yariv_1989_QuantumElectronics_p106-107}
Amnon Yariv, \textsl{Quantum Electronics} (John Wiley, New York, 1989), pp. 106--107.

\bibitem{Jancewicz_1988_MultivectorsandCliffordAlgebrainElectrodynamics_p32}
Bernard Jancewicz, \textsl{Multivectors and Clifford Algebra in Electrodynamics} (World Scientific, Singapore, 1988), p. 32.

\bibitem{Hestenes_2003_ajp71i2pp104-121_p110}
David Hestenes, ``Oersted medal lecture 2002: Reforming the mathematical language of physics,'' Am. J. Phys. \textbf{71}(2), 104--121 (2003).  See p. 110.

\bibitem{SugonFernandezMcNamara_2008_arXiv0811.3680v1_p9}
Quirino M. Sugon Jr., Carlo B. Fernandez, and Daniel J. McNamara, ``A geometric algebra reformulation of $2\times 2$ matrices: the dihedral group $\mathcal D_4$ in bra-ket notation,'' 	arXiv:0811.3680v1 [math-ph] (24 Nov 2008).  See p. 9.

\bibitem{Symon_1971_Mechanics_p408}
Keith R. Symon, \textsl{Mechanics} (Addison-Wesley, Reading, MA, 1971), $3^{rd}$ ed., p. 408.

\bibitem{Campbell_1994_JOSAA11i22pp618-622}
Charles E. Campbell, ``Ray vector fields,'' J. Opt. Soc. Am. A \textbf{11}(2), 618--622 (1994).

\bibitem{Harris_1997_OptVisSciv74i6pp349-366_p357}
W. F. Harris, ``Dioptric power: its nature and its representation in three- and four-dimensional space,'' Opt. Vis. Sci., \textbf{74}(6), 349--366 (1997).  See p. 357.

\bibitem{Sakurai_1967_AdvancedQuantumMechanics_p28}
Jun John Sakurai, \textsl{Advanced Quantum Mechanics} (Addison-Wesley, Reading, MA, 1967), p. 28.

\bibitem{LeBellac_1991_QuantumandStatisticalFieldTheory_p404}
Michel Le Bellac, \textsl{Quantum and Statistical Field Theory} (Clarendon, Oxford, 1991), p. 404.

\bibitem{KleinFurtak_1986_Optics_p152}
Miles V. Klein and Thomas E. Furtak, \textsl{Optics}, (John Wiley, New York, 1986), $2^{nd}$ ed., p. 152.

\bibitem{Meyer-Arendt_1984_IntroductiontoClassicalandModernOptics_p55}
Jurgen R. Meyer-Arendt, \textsl{Introduction to Classical and Modern Optics}, 2nd ed. (Prentice-Hall, Englewood Cliffs, NJ, 1984), p. 55.

\bibitem{BornWolf_1964_PrinciplesofOptics_p165}
Max Born and Emil Wolf, \textsl{Principles of Optics: Electromagnetic Theory of Propagation, Interference and Diffraction of Light} (Pergamon, Oxford, 1964), p. 165.

\bibitem{Welford_1974_AberrationsoftheOpticalSystem_pp22-23}
W. T. Welford, \textsl{Aberrations of the Optical System} (Academic, London, 1974), pp. 22--23.

\bibitem{SugonFernandezMcNamara_2008_arXiv0811.3680v1_p8}
See Ref. \cite{SugonFernandezMcNamara_2008_arXiv0811.3680v1_p9}, p. 8.

\bibitem{SugonMcNamara_2004_ajp72i1pp92-97_p95}
Quirino M. Sugon Jr. and Daniel J. McNamara, ``A geometric algebra reformulation of geometric optics,'' Am. J. Phys. \textbf{72}(1), 92-97.  See p. 95.  The concavity sign function $c_{\sigma\eta}$ is defined as the normalized dot product of the incident ray $\bm\sigma$ and the normal vector $\bm\eta$.

\bibitem{SugonMcNamara_2006_AIEP139pp179-224_pp206}
Quirino M. Sugon Jr. and Daniel J. McNamara, ''Ray tracing in spherical interfaces using geometric algebra,'' in \textsl{Advances in Imaging and Electron Physics} \textbf{139}, 179--224.  See pp. 206.  The concavity function $c_{\sigma\eta}$ is generalized to the relative direction sign function $c_{vw}$ for two vectors $\mathbf v$ and $\mathbf w$.

\newpage

\bibitem{SugonMcNamara_2008_arXiv:0810.5224v1_p3}
Quirino M. Sugon Jr. and Daniel J. McNamara, ``Paraxial meridional ray tracing equations from the unified reflection-refraction law via geometric algebra,'' 	arXiv:0810.5224v1 [physics.optics] (29 Oct 2008).  See p.~3.



\bibitem{Hecht_1987_Optics_p217}
Eugene Hecht, \textsl{Optics}, 2nd ed., with contributions by Alfred Zajac (Addison Wesley, Reading, MA, 1987), p. 217.

\bibitem{SugonMcNamara_2008_arXiv:0810.5224v1_p4}
See Ref. \cite{SugonMcNamara_2008_arXiv:0810.5224v1_p3}, p. 4.

\bibitem{SimonWolf_2000_josav17i2pp342-354_pp345-347}
R. Simon and Kurt Bernardo Wolf, ``Structure of the set of paraxial optical systems,'' J. Opt. Soc. Am. \textbf{17}(2), 342-354 (2000).  See pp. 345-347.

\bibitem{Hecht_1987_Optics_p219}
See Ref. \cite{Hecht_1987_Optics_p217}, p. 219.

\bibitem{BlakerRosenblum_1993_Optics_p37}
J. Warren Blaker and William M. Rosenblum, \textsl{Optics: An Introduction for Students of Engineering} (MacMillan, New York, 1993), p. 37.

\bibitem{KleinFurtak_1986_Optics_p156}
See Ref. \cite{KleinFurtak_1986_Optics_p152}, p. 156.

\bibitem{KleinFurtak_1986_Optics_p157}
See Ref. \cite{KleinFurtak_1986_Optics_p152}, p. 157.

\bibitem{SimmonsGuttmann_1970_StatesWavesandPhotons_p8}
Joseph W. Simmons and Mark J. Guttmann, \textsl{States, Waves and Photons: A Modern Introduction to Light} (Addison-Wesley, Reading, MA, 1970), p. 8.

\bibitem{Ghatak_1977_Optics_p52}
Ajoy Ghatak, \textsl{Optics} (Tata, New Delhi, 1977), p. 52.

\bibitem{NussbaumPhillips_1976_ContemporaryOpticsforScientistsandEngineers_p52}
See Ref. \cite{NussbaumPhillips_1976_ContemporaryOpticsforScientistsandEngineers_p10}, p. 52.

\bibitem{Goodman_1995_HandbookofOpticsI_ch1p71}
Douglas S. Goodman, ``General Principles of Geometric Optics,'' in \textsl{Handbook of Optics}, chapter 1 of vol. 1, \textsl{Fundamentals, Techniques, and Design} (McGraw-Hill, New York, 1995), p. 71.

\bibitem{NazarathyShamir_1982_josav72i3pp356-364_p362}
Moshe Nazarathy and Joseph Shamir, ``First-order optics---a canonical operator representation: lossless systems,'' J. Opt. Soc. Am. \textbf{72}(3), 356--364 (1982).  See p. 362.

\bibitem{SimmonsGuttmann_1970_StatesWavesandPhotons_p11}
See Ref. \cite{SimmonsGuttmann_1970_StatesWavesandPhotons_p8}, p. 11.

\bibitem{SimmonsGuttmann_1970_StatesWavesandPhotons_p10}
See Ref. \cite{SimmonsGuttmann_1970_StatesWavesandPhotons_p8}, p. 10.

\bibitem{NussbaumPhillips_1976_ContemporaryOpticsforScientistsandEngineers_p16}
See Ref. \cite{NussbaumPhillips_1976_ContemporaryOpticsforScientistsandEngineers_p10}, p. 16.

\bibitem{Ghatak_1977_Optics_p56}
See Ref. \cite{Ghatak_1977_Optics_p52}, p. 56.

\bibitem{SimmonsGuttmann_1970_StatesWavesandPhotons_p12}
See Ref. \cite{SimmonsGuttmann_1970_StatesWavesandPhotons_p8}, p. 12.

\bibitem{Goodman_1995_HandbookofOpticsI_ch1p45}
See Ref. \cite{Goodman_1995_HandbookofOpticsI_ch1p71}, p. 45.

\bibitem{Meyer-Arendt_1984_IntroductiontoClassicalandModernOptics_p48}
See Ref. \cite{Meyer-Arendt_1984_IntroductiontoClassicalandModernOptics_p55}, p. 48.
\end{thebibliography}
\end{document}